\documentclass[superscriptaddress, two column,
amsmath,
amssymb,
aps,
prl
]{revtex4-2}
\usepackage[%
colorlinks=true,
urlcolor=blue,
linkcolor=blue,
citecolor=blue
]{hyperref}
\usepackage{graphicx} 
\usepackage{setspace}

\begin{document}

\title{Global phase diagram of doped quantum spin liquid on the Kagome lattice}
\author{Zheng-Tao Xu}
\affiliation{State Key Laboratory of Low-Dimensional Quantum and Department of Physics, Tsinghua University, Beijing 100084, China}
\author{Zheng-Cheng Gu}
\email{zcgu@phy.cuhk.edu.hk}
\affiliation{Department of Physics, The Chinese University of Hong Kong, Shatin, New Territories, Hong Kong, China}
\author{Shuo Yang}
\email{shuoyang@tsinghua.edu.cn}
\affiliation{State Key Laboratory of Low-Dimensional Quantum and Department of Physics, Tsinghua University, Beijing 100084, China}
\affiliation{Frontier Science Center for Quantum Information, Beijing 100084, China}
\affiliation{Hefei National Laboratory, Hefei 230088, China}
\begin{abstract}
It has long been believed that doped quantum spin liquids (QSLs) can give rise to fascinating quantum phases, including the possibility of high-temperature superconductivity (SC) as proposed by P. W. Anderson's resonating valence bond (RVB) scenario.
The Kagome lattice $t$-$J$ model is known to exhibit spin liquid behavior at half-filling, making it an ideal system for studying the properties of doped QSL. 
In this study, we employ the fermionic projected entangled simplex state (PESS) method to investigate the ground state properties of the Kagome lattice $t$-$J$ model with $t/J = 3.0$. 
Our results reveal a phase transition from charge density wave (CDW) states to uniform states around a critical doping level $\delta_c \approx 0.27$. 
Within the CDW phase, we observe different types of Wigner crystal (WC) formulated by doped holes that are energetically favored. 
As we enter the uniform phase, a non-Fermi liquid (NFL) state emerges within the doping range $0.27 < \delta < 0.32$, characterized by an exponential decay of all correlation functions. 
With further hole doping, we discover the appearance of a pair density wave (PDW) state within a narrow doping region $0.32 < \delta < 1/3$. 
We also discuss the potential experimental implications of our findings.
\end{abstract}

\maketitle

\textit{Introduction} --- Quantum spin liquids (QSLs)  \cite{anderson1973resonating,balents2010spin, RevModPhys.89.025003, savary2016quantum, broholm2020quantum} arise from strong quantum fluctuations and exhibit fascinating quantum behaviors such as fractional excitations and long-range entanglement.
With its frustrated geometry, the Kagome lattice naturally gives rise to QSL in the spin-$1/2$ antiferromagnetic Heisenberg model. 
However, the exact physical nature of the QSL, e.g., whether it is a gapless $U(1)$ QSL \cite{PhysRevLett.98.117205, PhysRevB.87.060405, PhysRevB.89.020407, PhysRevLett.118.137202, PhysRevX.7.031020} or a gapped $Z_2$ QSL \cite{PhysRevLett.101.117203, Simeng2011, PhysRevLett.109.067201, PhysRevB.91.075112, PhysRevB.95.235107}, is still under debate. 
Furthermore, chiral QSL, which breaks the time-reversal symmetry, has also been discovered in certain extended Heisenberg models on the Kagome lattice \cite{PhysRevLett.108.207204, PhysRevLett.112.137202, Bauer2014, Gong2014, PhysRevB.91.041124, PhysRevB.92.094433}. 

Whether doped QSL can induce superconductivity has generated significant interest in condensed matter theory. 
The concept of QSL has been closely linked to the mechanisms of high-temperature superconductivity, particularly with Anderson's proposal of the resonating valence bond (RVB) state as a precursor to superconductivity in cuprates \cite{anderson1987resonating}. 
Extensive research efforts have been dedicated to investigating doped QSLs \cite{PhysRevB.35.8865, PhysRevLett.61.2376, PhysRevLett.60.2677, PhysRevLett.76.503, PhysRevB.71.174515, PhysRevLett.127.097002}. 
Among these studies, the Kagome lattice $t$-$J$ model has emerged as the simplest model to explore the properties of hole-doped spin liquids \cite{PhysRevLett.92.236404, PhysRevB.84.174409, PhysRevLett.111.097204, PhysRevLett.119.067002, peng2021doping, PhysRevLett.127.187003}. 
Large-scale density matrix renormalization group (DMRG) simulations have shown that doping a QSL tends to result in an insulating state characterized by the long-range charge density wave (CDW) order \cite{PhysRevLett.119.067002,peng2021doping}. 
The specific lattice geometry and doping concentration influence the emergent patterns, which can manifest as unidirectional stripe crystals or two-dimensional Wigner crystals (WCs). 
Furthermore, a comprehensive variational Monte Carlo (VMC) simulation, utilizing variational states parameterized by $SU(2)$ gauge rotation angles, has suggested the presence of a non-centrosymmetric chiral nematic superconducting state in a $2\times 2$ unit cell \cite{PhysRevLett.127.187003}.

\begin{figure}[tbp]
    \centering
    \includegraphics[width=1.0\columnwidth]{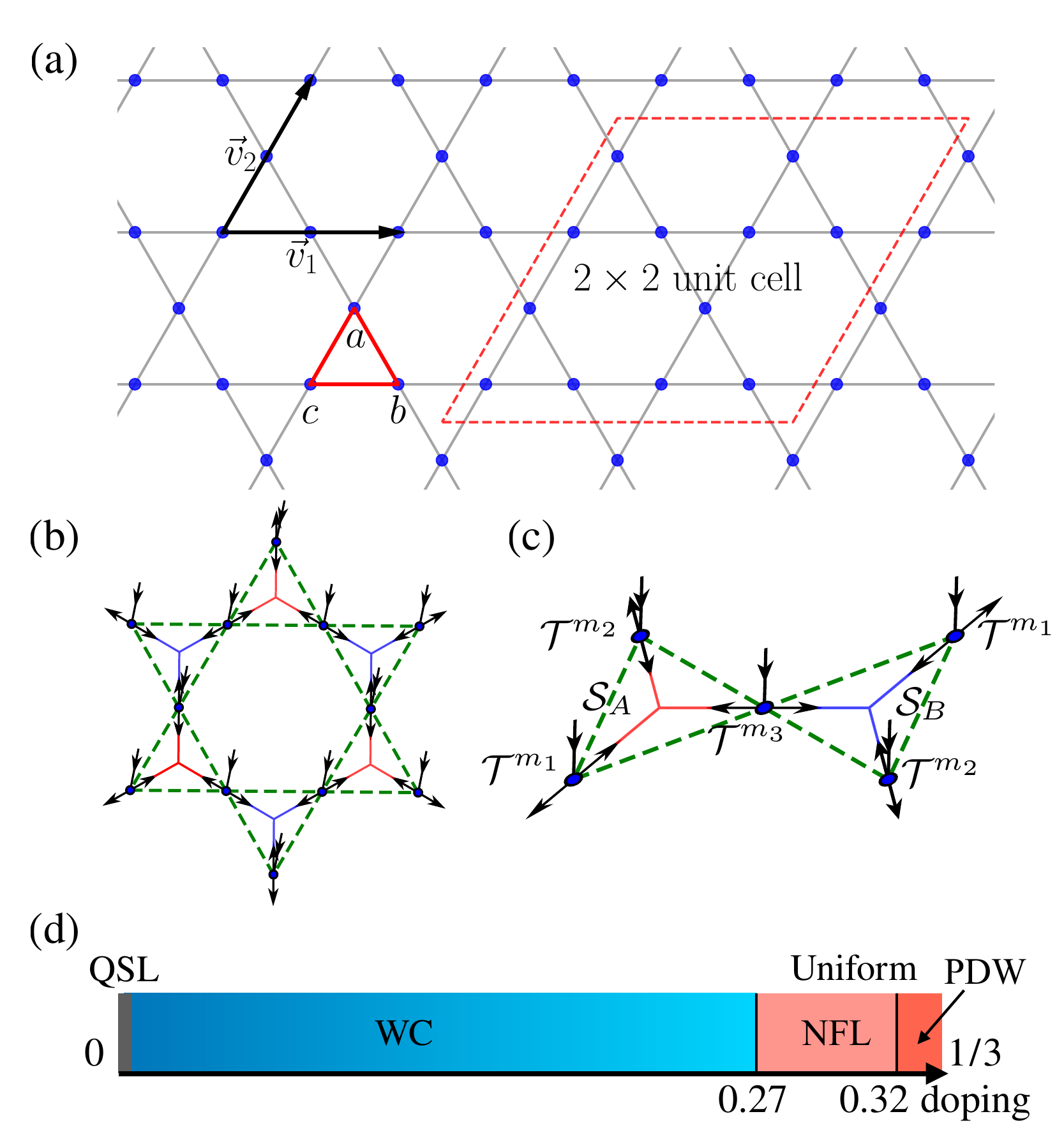} 
    \caption{
        (a) The Kagome lattice in the thermodynamic limit.
        The red dashed parallelogram denotes the $L_x \times L_y = 2 \times 2$ unit cell, and the arrows $\vec{v}_1$ and $\vec{v}_2$ show the primitive vectors.
        Each sublattice has three sites: $a$, $b$, and $c$.
        (b) Geometric structure of the Kagome lattice (dotted lines) and the tensor network state in the fermionic PESS representation \cite{PhysRevX.4.011025} (solid lines).
        (c) The $1\times 1$ unit cell of the Kagome lattice, compatible with the local tensors $\mathcal{T}^{m_a}, \mathcal{T}^{m_b}, \mathcal{T}^{m_c}, \mathcal{S}_A, \mathcal{S}_B$ with fermionic parity symmetry.
        (d) Phase diagram for the $t$-$J$ model with $t/J=3.0$ as a function of hole doping $\delta$ from $0$ to $1/3$.
        As hole doping increases, there is a phase transition from insulating WCs to uniform states at $\delta_c \approx 0.27$.
        The uniform PDW phase appears in a narrow region near $1/3$ doping.
    }
    \label{fig:1}
\end{figure}

In this Letter, we investigate the properties of the Kagome lattice $t$-$J$ model with $t/J = 3.0$ and hole doping ranging from $\delta = 0$ to $\delta = 1/3$ in the thermodynamic limit using a fermionic tensor network approach \cite{gu2010grassmann, PhysRevB.88.115139, PhysRevB.95.075108, Bultinck_2017}. 
Figure \ref{fig:1}(d) represents the phase diagram summarizing our findings. 
An intriguing observation is the phase transition from CDW to uniform states, which occurs at a critical doping level $\delta_c \approx 0.27$. 
Within the CDW phase, we find that WCs formed by doped holes \cite{PhysRevLett.119.067002} are energetically favored over stripe states in the lightly doped region. 
As the hole doping increases to $\delta \simeq 0.27$, the CDW states maintain the characteristics of the WCs. 
In particular, we discover a nearly degenerate trihexagonal (TrH) CDW state at $\delta=1/6$, which adds an attractive feature to the phase diagram. 
Moving into the uniform phase, a significant finding is the ground state that exhibits pair density wave (PDW) pairing symmetry in a narrow doping range $0.32 < \delta < 1/3$. 
However, within the doping range $0.27 < \delta < 0.32$, the uniform states do not exhibit any superconducting order, and we conclude that it could be a non-Fermi liquid (NFL) with exponential decay of single-particle and spin-spin correlation functions.
    
\textit{Model and Method} --- 
The Hamiltonian of the $t$-$J$ model on the Kagome lattice reads
\begin{equation}
H = -t\sum_{\langle ij\rangle , \sigma}\left( \tilde{c}_{i,\sigma}^\dagger \tilde{c}_{j,\sigma} + h.c.\right) + J\sum_{\langle i, j\rangle} \left( \vec{S}_i\cdot\vec{S}_j - \frac{1}{4}\hat{n}_i \hat{n}_j\right), \nonumber
\end{equation}
where $\tilde{c}_{i,\sigma} = \hat{c}_{i\sigma}(1-\hat{n}_{i\bar{\sigma}})$ is the electron annihilation operator with spin $\sigma = \{\uparrow, \downarrow\}$ in the no-double-occupancy subspace on site $i$.
$\vec{S}_i$ is the spin-$1/2$ operator, and $\hat{n}_{i} = \sum_{\sigma} \hat{c}_{i,\sigma}^\dagger \hat{c}_{i,\sigma}$ denotes the local electron density.
We consider the parameter $t/J = 3.0$ and control hole doping $\delta$ by varying the chemical potential $\mu$.

We use the simple update (SU) method based on the imaginary time evolution technique \cite{PhysRevLett.98.070201, PhysRevX.4.011025} to obtain ground state wave functions.
We choose an appropriate imaginary time step $\Delta \tau$, which gradually decreases from $\Delta \tau_{\mathrm{start}} = 10^{-2}$ to $\Delta \tau_{\mathrm{start}} = 10^{-5}$, to ensure the convergence and efficiency of optimization.
At the end of SU, the average change in Schmidt weight is less than $10^{-9}$.
We then calculate the physical measurements using the variational uniform matrix product state (VUMPS) algorithm \cite{PhysRevB.98.235148, PhysRevB.97.045145, 10.21468/SciPostPhysLectNotes.7, PhysRevB.108.035144}.
We use effective environmental tensors to calculate related quantities by contraction, such as variational energy, local charge density, superconductivity, etc.
The environmental bond dimension $\chi$ influences the accuracy of the physical quantities.
With a sufficient $\chi \gtrsim 4 D$, the relative errors for these physical quantities are of the order $10^{-4}$.

\begin{figure}[tbp]
    \centering
    \includegraphics[width=1.0\columnwidth]{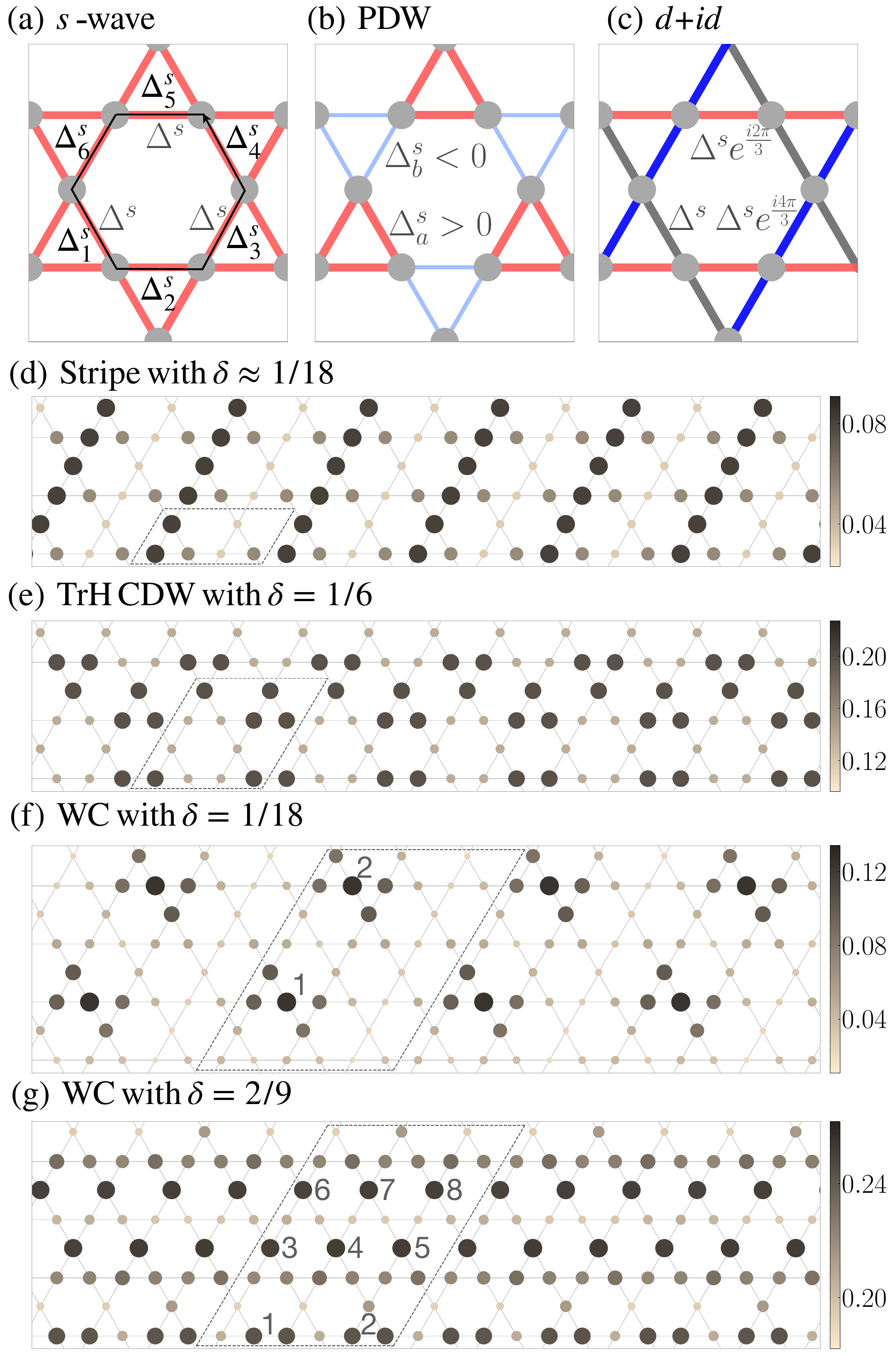} 
    \caption{
        Three uniform states with different pairing symmetries, including (a) the $s$-wave state, (b) the PDW state, and (c) the $d+id$-wave state.
        (d-g) Patterns of CDW states on various $L_x\times L_y$ unit cells with bond dimension $D=16$.
        The diameter (greyscale) of the disk in each pattern scales with the local hole density.
        The dashed parallelogram indicates a $L_x\times L_y$ unit cell.
        (d) The stripe state on a $2\times 1$ unit cell at $\delta \approx 1/18$.
        (e) The TrH CDW state on a $2 \times 2$ unit cell at $\delta = 1/6$.
        (f) The Wigner crystal on a $3 \times 4$ unit cell at $\delta = 1/18$.
        (g) The Wigner crystal on a $3 \times 4$ unit cell at $\delta = 2/9$.
    }
    \label{fig:2}
\end{figure}

\begin{figure}[tbp]
    \centering
    \includegraphics[width=1.0\columnwidth]{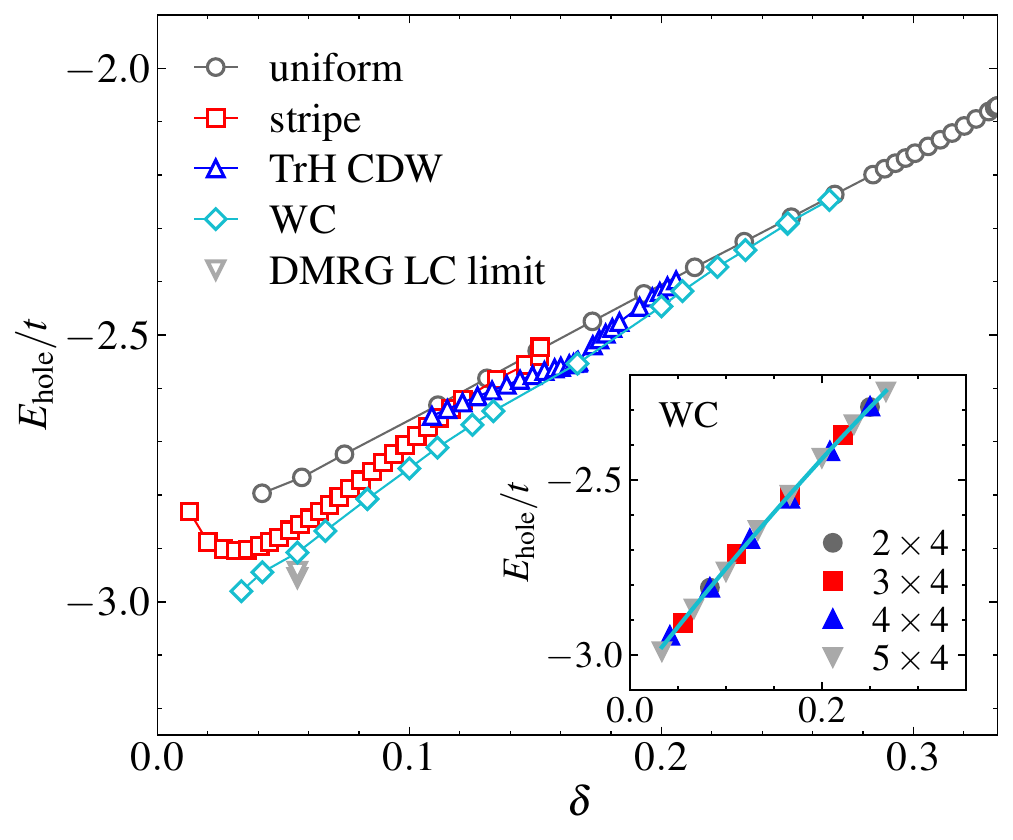} 
    \caption{
        Energies per hole $E_{\mathrm{hole}}$ of various competing states for $t/J = 3.0$ and $D = 16$ as a function of hole doping $\delta$.
        The inset shows WCs on $L_x \times 4$ $(L_x = 2,3,4,5)$ unit cells, which have nearly the same energies per hole.
        The DMRG results show the stripe state with $E_{\mathrm{hole}} = -2.955$ on the YC-6 cylinder and the holon WC with $E_{\mathrm{hole}} = -2.943$ on the YC-8 cylinder at $\delta = 1/18$ in the long cylinder (LC) limit \cite{PhysRevLett.119.067002}.
    }
    \label{fig:3}
\end{figure}

\textit{Global Phase Diagram} ---
Depending on the size of the unit cell and the initial ansatz, the converged state can exhibit either uniform or CDW behavior. 
Figure \ref{fig:2} shows all relevant competing states in different unit cells. 
In the following, we will first discuss the physical properties of all these states and then determine the true ground state by comparing their energies.

We start with a uniform ansatz in a $1 \times 1$ unit cell.
In the thermodynamic limit, the uniform state always breaks the charge $U(1)$ symmetry spontaneously, and we can detect the SC order in the spin-singlet channel in the real space using $\Delta^s_{ij} = \frac{1}{\sqrt{2}} \langle \hat{c}_{i\uparrow} \hat{c}_{j\downarrow} - \hat{c}_{i\downarrow} \hat{c}_{j\uparrow}\rangle$.
The possible pairing states in the Kagome lattice are classified by irreducible representations of the $C_{6v}$ point group symmetry \cite{PhysRevB.105.075118}.
Figures \ref{fig:2}(a), \ref{fig:2}(b), and \ref{fig:2}(c) show three distinct uniform states in large hole doping, with different pairing symmetries $\vec\Delta^s = (\Delta^s_1,\Delta^s_2,\Delta^s_3,\Delta^s_4,\Delta^s_5,\Delta^s_6)$.
These are 1) the (extended) $s$-wave state with $\vec\Delta^s = \Delta^s(1,1,1,1,1,1)$, 2) the PDW state with $\vec\Delta^s = (\Delta_a^s,\Delta_b^s,\Delta_a^s,\Delta_b^s,\Delta_a^s,\Delta_b^s)$, where $\Delta^a > 0$ and $\Delta^b < 0$, and 3) the chiral $d+id$-wave state with $\vec\Delta^s = \Delta^s(1,e^{i2\pi/3},e^{i4\pi/3},1,e^{i2\pi/3},e^{i4\pi/3})$.
The PDW state has pairing orders with sign oscillations along all three directions.

For the non-uniform ansatz in $L_x \times L_y$ unit cells, we find various CDW states for $\delta < 0.27$.
Figures \ref{fig:2}(d), \ref{fig:2}(e), \ref{fig:2}(f), and \ref{fig:2}(g) show the characteristics of three types of different CDW states: stripe state, TrH CDW state, and WC. Figure \ref{fig:2}(d) shows a typical stripe state in a $2\times 1$ unit cell. 
This wave function breaks the horizontal translational symmetry in $L_x \times 1$ ($L_x \geq 2$) unit cells, resulting in a modulation of the charge density in the $\vec{v}_1$ direction. 
The stripe periodicity depends on the unit cell size.
DMRG simulations also report the stripe crystal in the YC-6 cylinder in the lightly doped region \cite{PhysRevLett.119.067002,peng2021doping}.
The stripe states compete with the SC order, and we do not find any SC order in the stripe states. 
Figure \ref{fig:2}(e) illustrates the TrH CDW state in a $2\times 2$ unit cell at $\delta = 1/6$. 
This state emerges mainly for $0.1 < \delta < 0.2$ and preserves the $C_3$ rotational symmetry.
In the TrH CDW state, holes tend to localize at the triangles of the Kagome lattice, whereas electrons localize at the hexagons. 
In the extrapolation as $D\rightarrow \infty$, the state does not have net magnetism at each lattice site.
Finally, we further examine the non-uniform ansatz in large $L_x\times L_y$ unit cells ($L_x L_y \geq 8$, $L_x \geq 2$, and $L_y \geq 2$) and find a series of WCs.
Each WC has an effective repulsive interaction between the doped holes.
Figure \ref{fig:2}(f) shows a WC in the $3\times 4$ unit cell at $\delta = 1/18$ (two doped holes per unit cell).
Each doped hole forms a localized cluster, and two holes repel each other in the two-dimensional real space, leading to crystallization.
Figure \ref{fig:2}(g) displays the WC in a $3\times 4$ unit cell at $\delta = 2/9$ (eight doped holes per unit cell).
The local hole density at the eight sites is higher than at the surrounding sites, indicating the state of doped holes with an effective repulsive interaction.
We also study various unit cells with $L_y=4$ and $L_x \in \{2,3,4,5\}$. 
Each unit cell supports the formation of these crystals (see Supplemental Materials for more details).

In Fig. \ref{fig:3}, we compare the energy of the uniform states and various CDW states for $D=16$.
We calculate the energy per hole defined as $E_{\mathrm{hole}}(\delta) = [E_0(\delta) - E_0(0)]/\delta$, where $E_0(\delta)$ is the energy per site, and $E_0(0) = -0.937526J$ at half filling is from a PESS calculation for the Heisenberg model on the Kagome lattice \cite{PhysRevLett.118.137202}.
In the inset of Fig. \ref{fig:3}, our calculations show that the energies of the WCs on different unit cells lie close to a line as a function of hole doping. 
We observe a phase transition from insulating WCs to uniform states at $\delta = 0.27$.
In the light hole doping regime $\delta < 0.16$, WCs have lower energies than stripe states in the thermodynamic limit. Large-scale DMRG simulations also show that the lightly doped QSL in the kagome lattice $t$-$J$ model leads to the WC of spinless holons \cite{PhysRevLett.119.067002,peng2021doping}.
We observe a similar absence of magnetizations for these WCs at $\delta < 0.13$ in the extrapolation with $1/D$ (see Supplemental Materials for more details).
This suggests the presence of spinless holons in these lightly doped QSLs. However, the repulsive interactions among holons are still robust, and the formation of uniform superconductivity will cost much higher energy at low doping.
Interestingly, we also observe a near degeneracy of the TrH CDW state and the WC at $\delta = 1/6$.
The TrH CDW state at $1/6$ doping can be roughly viewed as a WC with two doped holes.
However, the TrH CDW states are not energetically favored at $\delta \neq 1/6$.
In addition, these insulating WCs also survive on the electron doping side with negative $t$ (see more details in Supplemental Materials).
All our results with $D=12$, $D=14$, and $D=16$ show a consistent trend in the phase transition.
The result with $ D=16 $ is reliable enough to stretch the global phase diagram.

\begin{figure}[tbp]
    \centering
    \includegraphics[width=1.0\columnwidth]{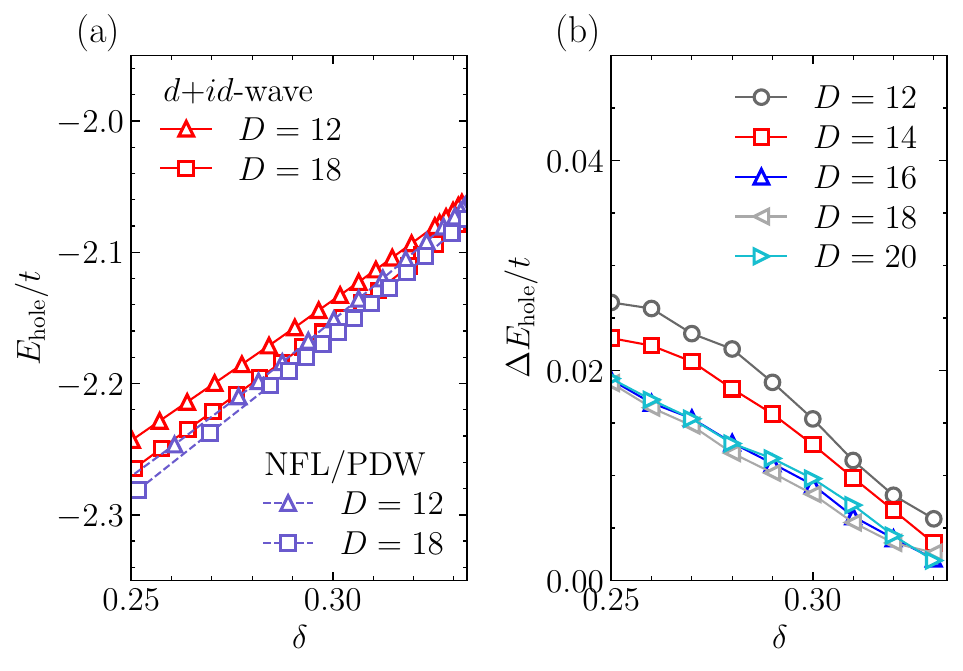} 
    \caption{
        (a) Energies per hole of various uniform states.
        The blue dashed lines show the NFL and PDW states.
        (b) Energy difference $\Delta E_{\mathrm{hole}}$ between the uniform $d$+$id$-wave states and the NFL/PDW states. 
    }
    \label{fig:4}
\end{figure}

\begin{figure}[tbp]
    \centering
    \includegraphics[width=1.0\columnwidth]{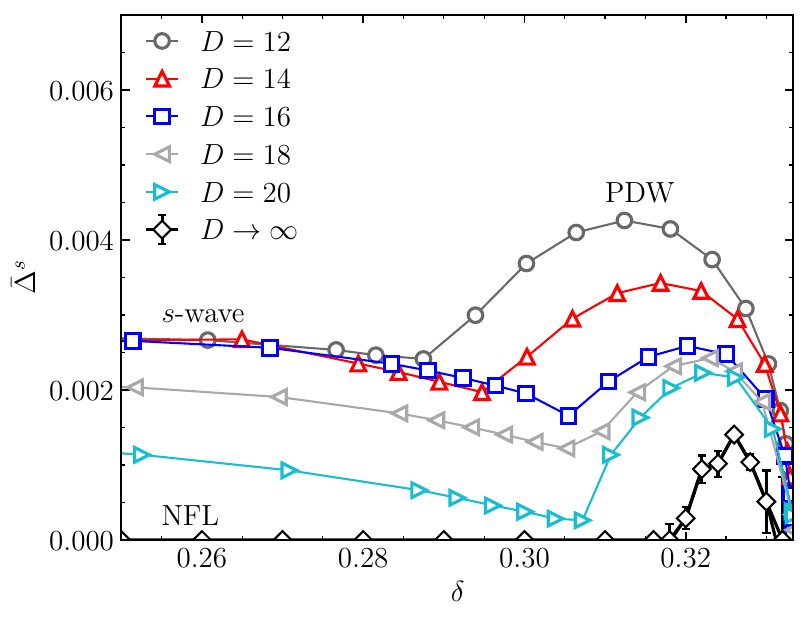} 
    \caption{
        The average SC order $\bar\Delta^s$ for the uniform states, including the NFL and PDW states.
        For each finite $D$, the discontinuity indicates a phase transition from the ``$s$-wave'' state to the PDW state.
        when $D \rightarrow \infty$, the SC order of PDW states remains at $\delta > 0.32$, but the $s$-wave does not.
        We call the latter the NFL.
    }
    \label{fig:5}
\end{figure}

\begin{figure}[tbp]
    \centering
    \includegraphics[width=1.0\columnwidth]{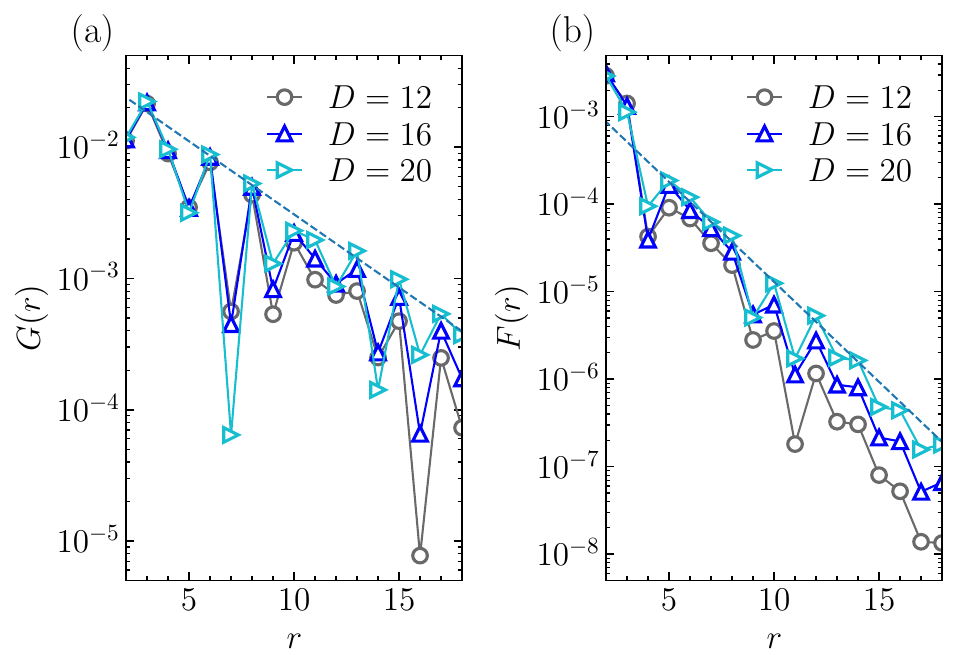} 
    \caption{
        Two correlation functions of the state with $\delta \approx 0.285$ along the $\vec{v}_1$ direction for the NFL phase. 
        (a) The single-particle Green's funciton $G(r)$.
        (b) The spin-spin correlation function $F(r)$.
        Both correlation functions are in semi-logarithmic scale, i.e., $G(r) \sim e^{-r/\xi_G}$ with $\xi_G \approx 3.9$ site spacing ($\approx 1.9$ lattice spacing) for $D=20$, and $F(r) \sim e^{-r/\xi_F}$ with $\xi_F \approx 1.9$ site spacing for $D=20$.
    }
    \label{fig:6}
\end{figure}

\textit{Non-fermi Liquid and Uniform PDW States} --- WCs are energetically unstable and merge into a uniform ground state when $ \delta>0.27 $. In this uniform phase, we find three distinct states with different SC pairing symmetries: ``$s$-wave'', PDW, and $d$+$id$-wave states. 
The ``$s$-wave'' state has the lowest energy within a narrow doping range $0.27<\delta<0.32$. 
Nevertheless, the SC order of the uniform ``$s$-wave'' states disappears in the $1/D$ extrapolation, implying the emergence of an NFL phase.
Surprisingly, the NFL transitions to a PDW state as hole doping increases to $\delta=0.32$ and survives in a tiny doping range $0.32<\delta<0.33$.
Although the uniform $d$+$id$-wave state exists for a broad doping range (see Supplemental Materials for details), it is not the lowest energy state. 

Figure \ref{fig:4} compares the hole energies of the NFL and PDW states with the $d+id$-wave states as a function of hole doping $\delta$.
To compare the energy more accurately, we calculate the energy difference $\Delta E_{\mathrm{hole}}$ in Fig. \ref{fig:4}(b), which is defined as $\Delta E_{\mathrm{hole}} = E_{\mathrm{hole}}^{d+id} - E_{\mathrm{hole}}^{\mathrm{NFL/PDW}}$.
Here, $E_{\mathrm{hole}}^{d+id}$ is the hole energy of the $d+id$-wave state, and $E_{\mathrm{hole}}^{\mathrm{NFL/PDW}}$ is the hole energy of the NFL or PDW state.
The NFL and PDW states have energies lower than the chiral states for each $D$.
As $D$ increases, $\Delta E_{\mathrm{hole}}$ no longer decreases for $D \geq 16$, indicating that the chiral state is not the energetically favored state. 
Figure \ref{fig:5} shows the SC order $\bar\Delta^s = \sum_{i=1}^6 |\Delta_i^s|/6$ for the NFL and PDW states, with the result of a linear extrapolation with $1/D$.
The SC order changes from the ``$s$-wave'' state to the PDW state as hole doping increases for each $D$.
The ``$s$-wave'' state has an initial SC order at each $D$, but it decays quickly when $D \geq 16$, eventually disappearing in the extrapolation as $D\rightarrow\infty$.
We conjecture that the uniform state could be an NFL for $0.27 < \delta < 0.32$, while the PDW state still has a persistent SC order for $0.32<\delta<0.33$ in the extrapolation. 
However, because the PDW exists only in a narrow range of doping, further studies with global optimization are necessary to confirm the existence of such an exotic phase.

We analyze additional correlation functions to confirm the presence of the NFL phase. 
Figure \ref{fig:6} displays two types of correlation functions for the NFL with $\delta \approx 0.285$: the single-particle Green's function $G(r)$ and the spin-spin correlation function $F(r)$. 
Both $G(r)$ and $F(r)$ exhibit exponential decay with short correlation lengths $\xi_G \approx 3.9$ and $\xi_F \approx 1.9$ for $D=20$. 
This behavior differs from the traditional Fermi liquid phase with power-law decay single-particle Green's function and magnetically ordered phases with finite magnetization. 
In comparison, we also examine these correlation functions in the WC phase. 
While the single-particle Green's function still exhibits exponential decay, the spin-spin correlations generally display power-law behavior (see Supplemental Materials for more details), which is significantly different from the NFL phase.

\textit{Discussion and Conclusion} ---
Using fermionic PESS, we study the ground state phase diagram of the Kagome lattice $t$-$J$ model with $t/J = 3.0$ as a function of hole doping $\delta \in [0,1/3]$.
We find a phase transition from CDW states to uniform states around $\delta = 0.27$.
In the CDW phase, WCs are the lowest energy states in large unit cells.
The TrH CDW state at $\delta =1/6$ is nearly degenerate with the WC.
In the uniform phase, the existence of uniform PDW states in the narrow doping region $0.32 < \delta < 1/3$ is of particular interest.
The uniform states are NFL for $0.27 < \delta < 0.32$. 
The PDW state, induced by a strong electronic correlation \cite{PhysRevLett.130.026001}, provides theoretical foundations for exploring unconventional phenomena in strongly correlated Kagome materials.

The crystallization of lightly doped spin liquids in the Kagome lattice has been attributed to the near-degeneracy between QSLs and valence-bond-crystalline phases. 
This implies that holon charge carriers form a crystal structure due to effective repulsive interactions between them, rather than a possible superconducting phase \cite{PhysRevLett.119.067002}. 
It is worth noting that the NFL phase is also observed in the honeycomb lattice $t$-$J$ model \cite{miao2023spincharge}, suggesting that it could be a common feature in correlated electronic systems. 
However, our results indicate that the geometric frustrations in the Kagome lattice $t$-$J$ model make a simple $s$-wave or $d+id$-wave uniform superconducting phase unstable under both hole doping and electron doping conditions.
Experimentally, the well-known materials herbertsmithite ZnCu$_3$(OH)$_6$Cl$_2$ \cite{doi:10.1021/ja053891p, Han2012, fu2015evidence} and the Zn-substituted barlowite Cu$_3$Zn(OH)$_6$FBr \cite{PhysRevLett.113.227203, Feng_2017, Smaha2020} are promising candidates for kagome antiferromagnetic systems. 
It would be exciting if they could be doped.

\textit{Acknowledgement}
We thank H.-C. Jiang for providing some data in Ref. \cite{PhysRevLett.119.067002} for comparison.
Z.T.X. and S.Y. are supported by the National Natural Science Foundation of China (NSFC) (Grant No. 12174214 and No. 92065205) and the Innovation Program for Quantum Science and Technology (Project 2021ZD0302100). 
Z.C.G. is supported by funding from Hong Kong's Research Grants Council (CRF C7012-21GF, RFS2324-4S02) and Direct Grant No. 4053578 from The Chinese University of Hong Kong.

\bibliography{refs}
\appendix

\section{CDW Phase}
\label{app:cdw}

\begin{figure}[tbp]
    \centering
    \includegraphics[width=1.0\columnwidth]{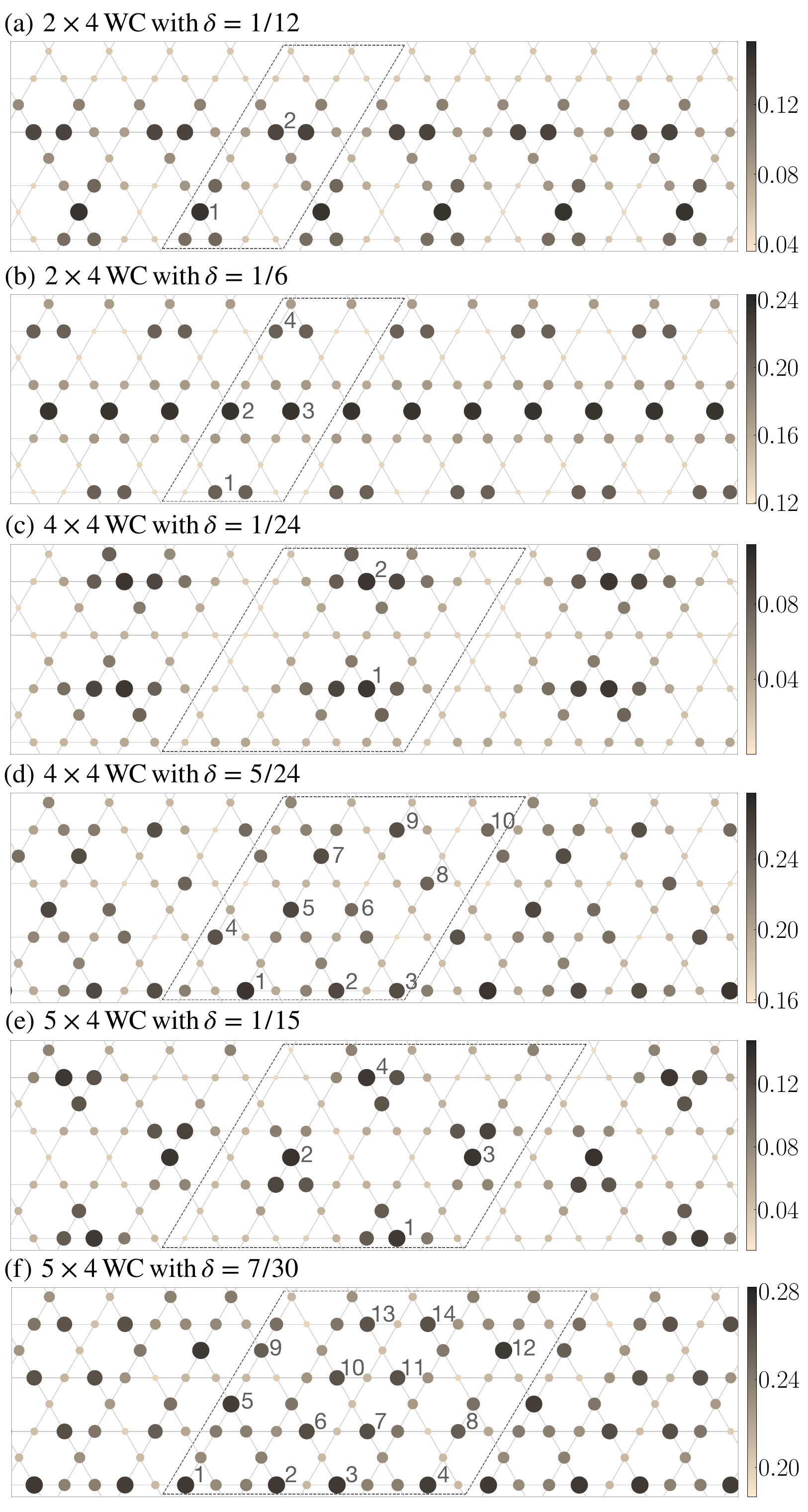} 
    \caption{
        Patterns of WCs with different hole doping $\delta$ on large unit cells.
        WCs in the $2\times 4$ unit cell with 
        (a) $\delta = 1/12$ ($2$ doped holes) and
        (b) $\delta = 1/6$ ($4$ doped holes).
        WCs in the $4\times 4$ unit cell with 
        (c) $\delta = 1/24$ ($2$ doped holes) and
        (d) $\delta = 5/24$ ($10$ doped holes).
        WCs in the $5\times 4$ unit cell with 
        (e) $\delta = 1/15$ ($4$ doped holes) and
        (f) $\delta = 7/30$ ($14$ doped holes).
        We mark the positions of the maximum local hole density by numbers.
        The number of doped holes equals the number of marked sites.
    }
    \label{fig:7}
\end{figure}

\begin{figure}[tbp]
    \centering
    \includegraphics[width=1.0\columnwidth]{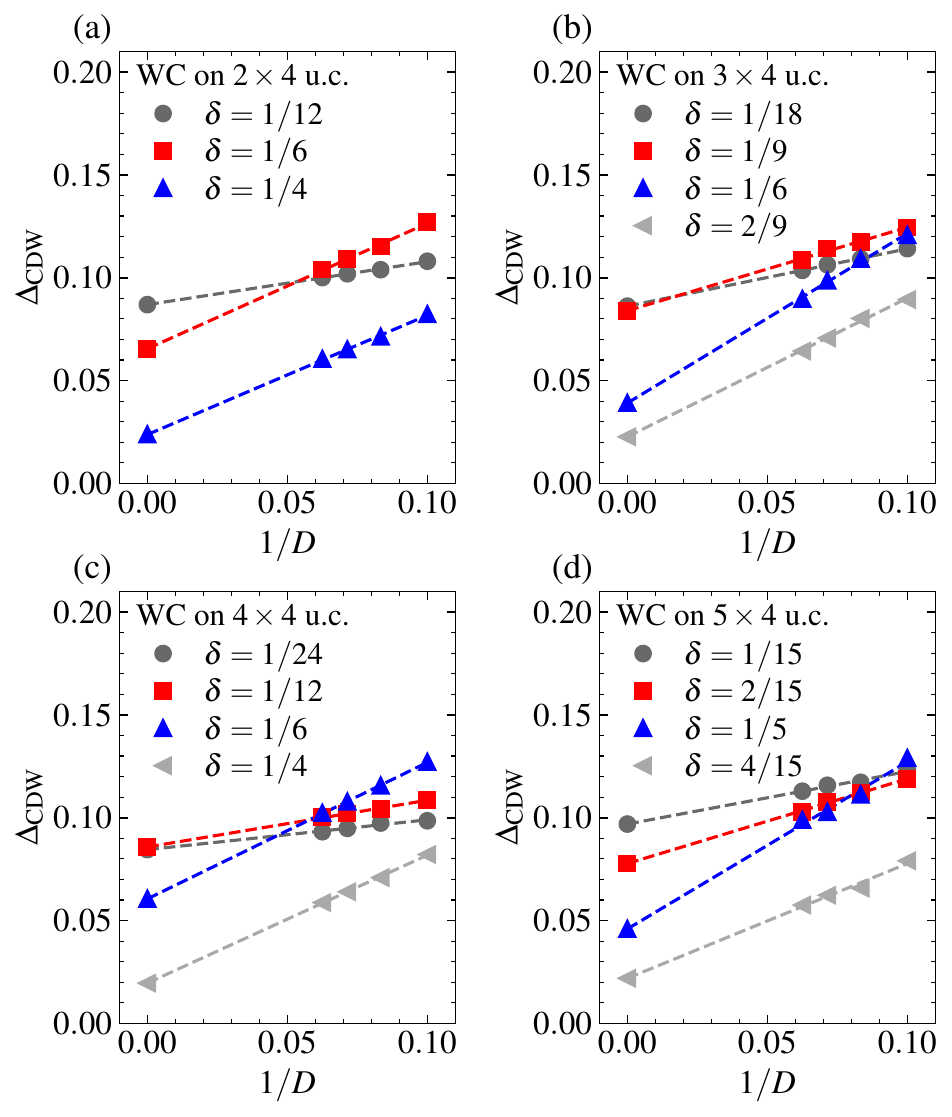} 
    \caption{
        The CDW order $\Delta_{\mathrm{CDW}}$ as a function of $1/D$ for WCs on 
        (a) $2 \times 4$, 
        (b) $3 \times 4$,
        (c) $4 \times 4$,
        and (d) $5 \times 4$ unit cells.
        The order parameter $\Delta_{\mathrm{CDW}} = \delta_{\mathrm{max}} - \delta_{\mathrm{min}}$, where $\delta_{\mathrm{max}}$ is the maximum local hole density, and $\delta_{\mathrm{min}}$ is the minimum local hole density.
    }
    \label{fig:8}
\end{figure}

\begin{figure}[tbp]
    \centering
    \includegraphics[width=1.0\columnwidth]{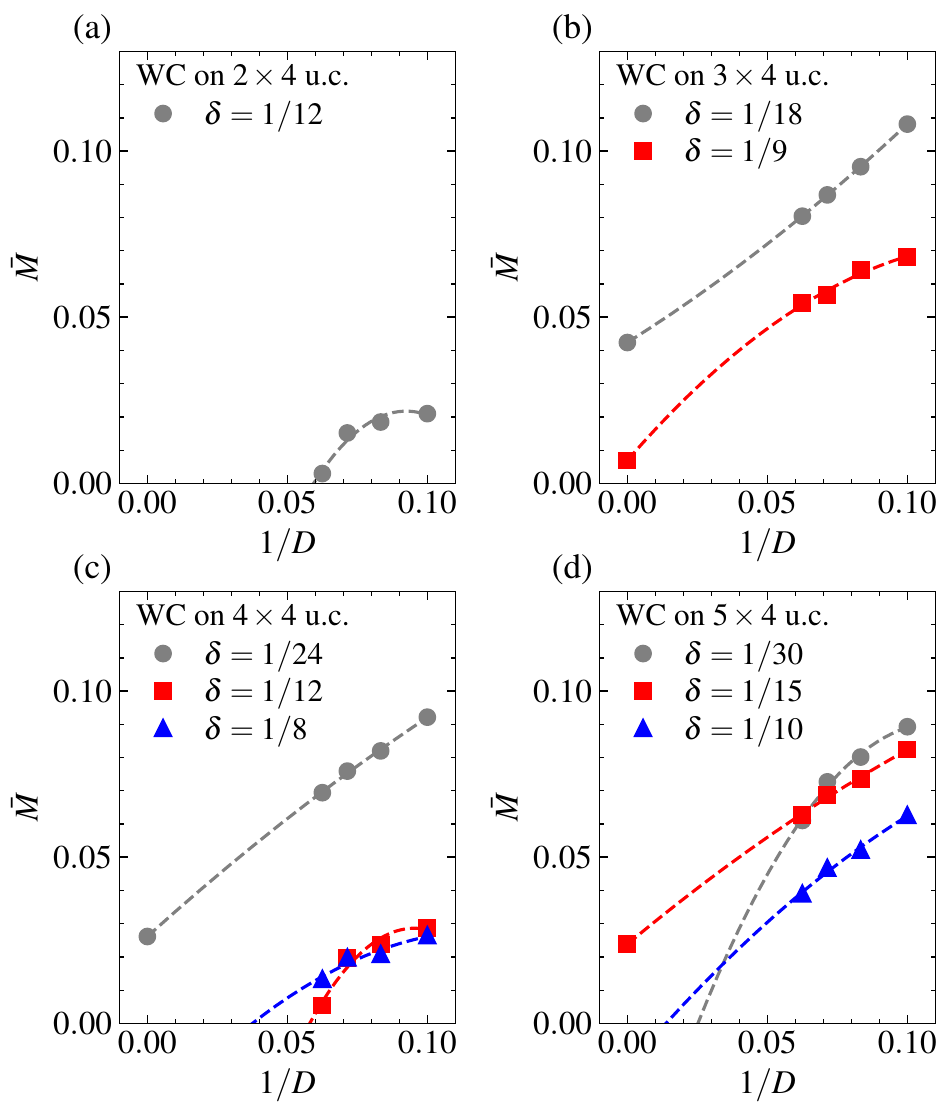} 
    \caption{
        The average magnetization $\bar{M}$ as a function of $1/D$ for WCs with light hole doping on 
        (a) $2 \times 4$, 
        (b) $3 \times 4$,
        (c) $4 \times 4$,
        and (d) $5 \times 4$ unit cells.
    }
    \label{fig:9}
\end{figure}

\begin{figure}[tbp]
    \centering
    \includegraphics[width=1.0\columnwidth]{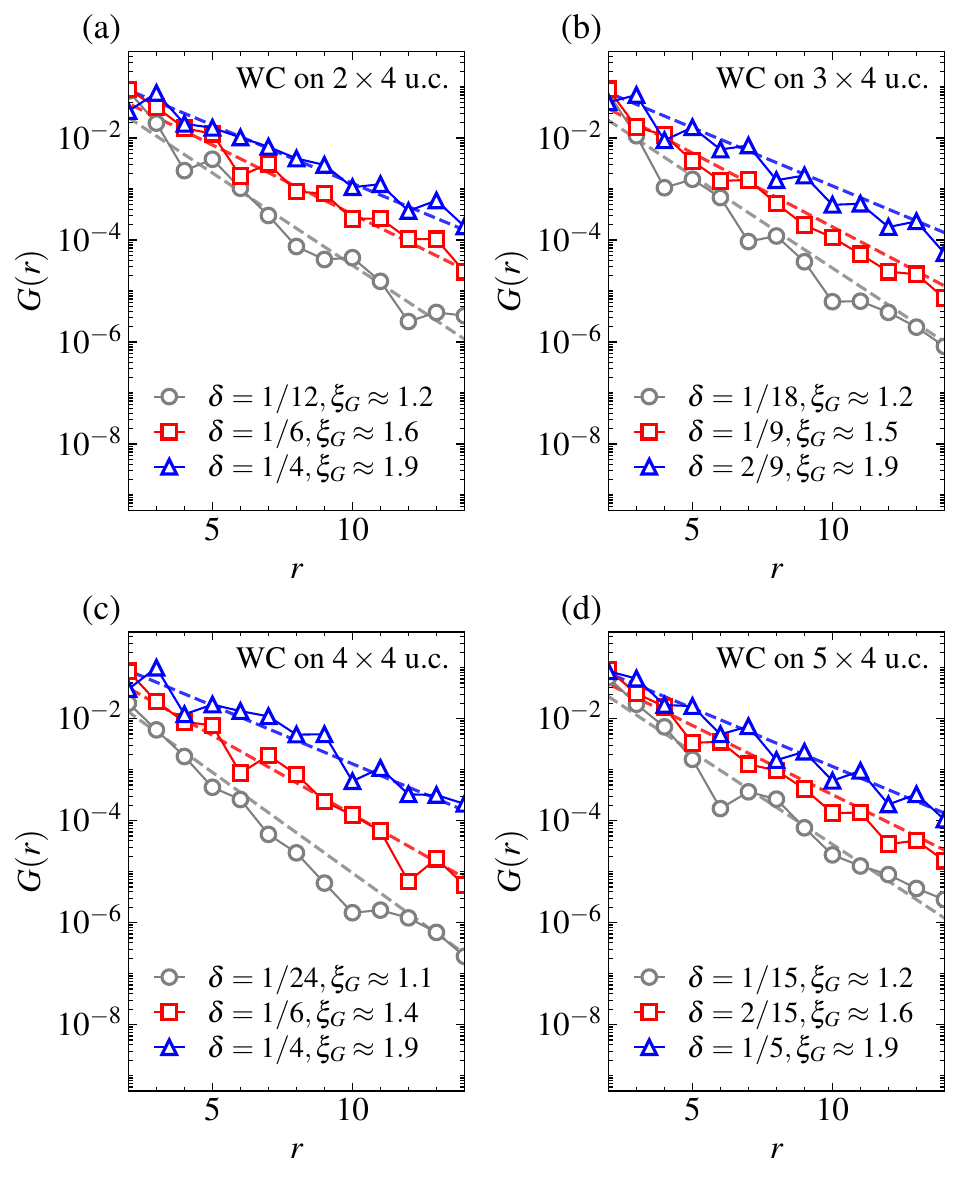} 
    \caption{
        The single-particle Green's function $G(r)$ along the $\vec{v}_1$ direction for WCs with $D=16$ on 
        (a) $2 \times 4$, 
        (b) $3 \times 4$,
        (c) $4 \times 4$,
        and (d) $5 \times 4$ unit cells.
        Here, $r$ is the site spacing between the $b$ and $c$ sites.
        All single-particle Green's functions are in semi-logarithmic scale, i.e., $G(r) \sim e^{-r/\xi_G}$.
    }
    \label{fig:10}
\end{figure}

\begin{figure}[tbp]
    \centering
    \includegraphics[width=1.0\columnwidth]{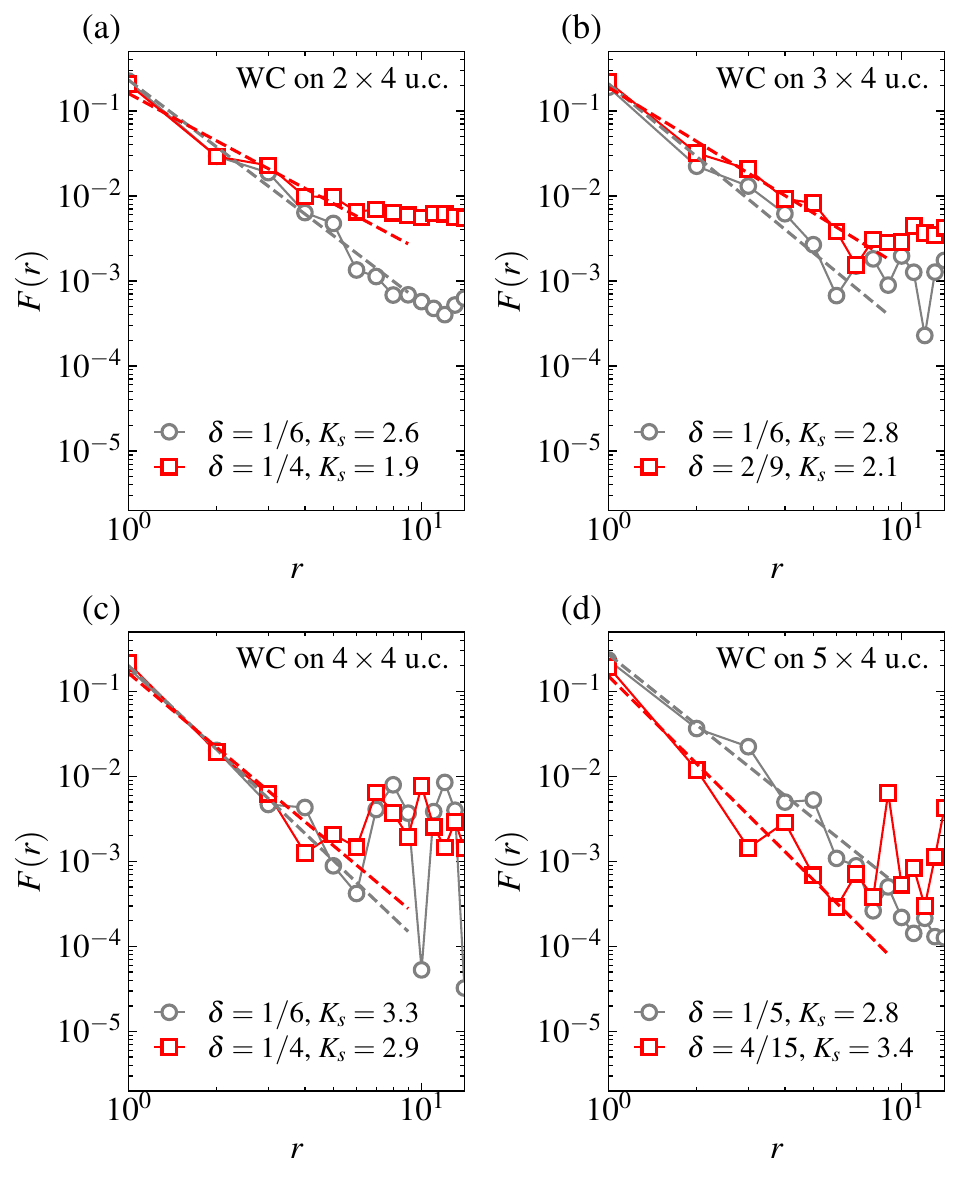} 
    \caption{
        The spin-spin correlation function $F(r)$ along the $\vec{v}_1$ direction for WCs with $D=16$ on 
        (a) $2 \times 4$, 
        (b) $3 \times 4$,
        (c) $4 \times 4$,
        and (d) $5 \times 4$ unit cells.
        Here, $r$ is the site spacing between $b$ and $c$ sites.
        $F(r)$ in double-logarithmic scale shows $F(r) \sim r^{-K_s}$ at the short-range distance.
    }
    \label{fig:11}
\end{figure}

\begin{figure}[tbp]
    \centering
    \includegraphics[width=1.0\columnwidth]{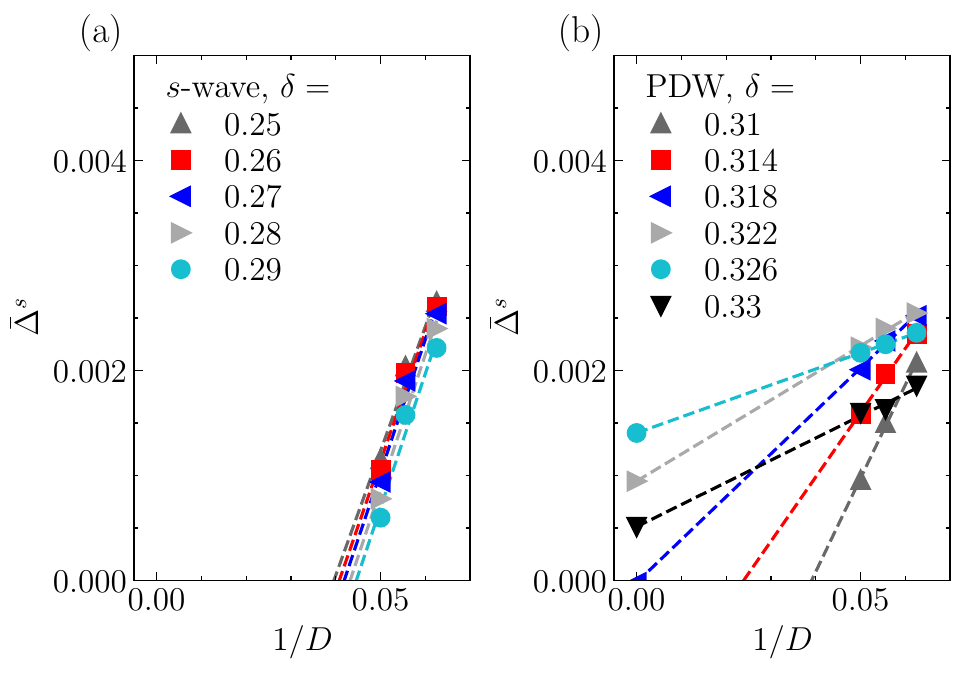} 
    \caption{
        The extrapolation details with $1/D$ for the SC order $\bar\Delta^s$ at different hole doping $\delta$, including uniform states with (a) $s$-wave and (b) PDW pairing symmetry.
    }
    \label{fig:12}
\end{figure}

\begin{figure}[tbp]
    \centering
    \includegraphics[width=1.0\columnwidth]{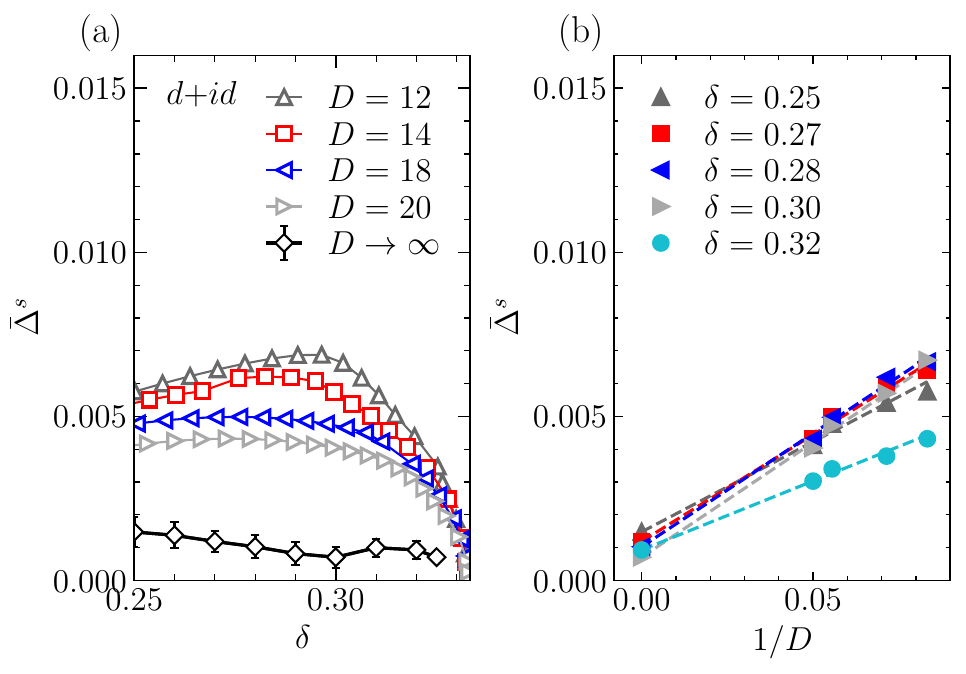} 
    \caption{
        (a) The average SC order $\bar\Delta^s$ for uniform $d$+$id$-wave states.
        (b) The extrapolation details with $1/D$ for $\bar\Delta^s$ at different hole doping $\delta$.
    }
    \label{fig:13}
\end{figure}

\begin{figure}[tbp]
    \centering
    \includegraphics[width=1.0\columnwidth]{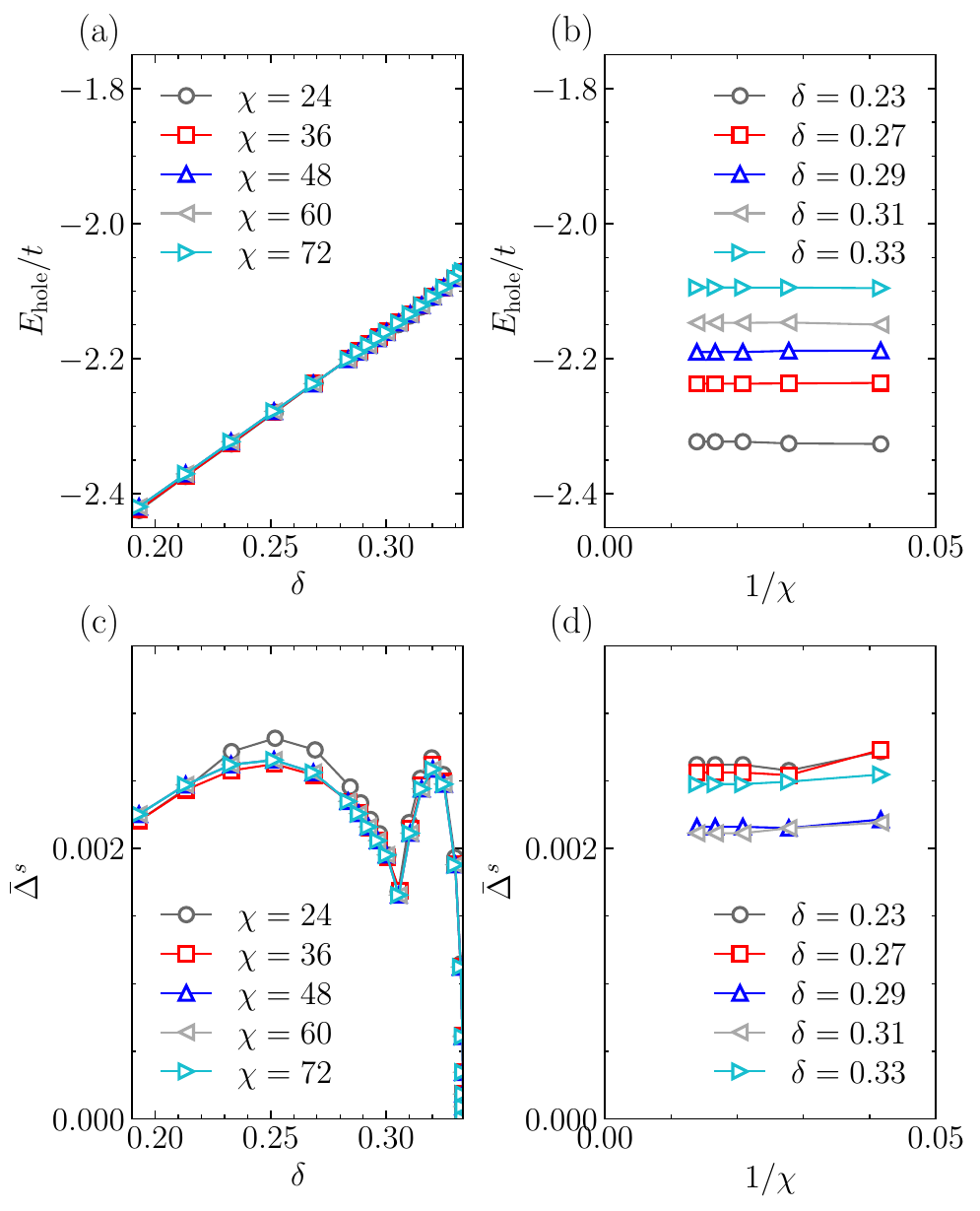} 
    \caption{
        Results for the NFL and PDW states with $D=16$ as the environmental bond dimension $\chi$ increases.
        (a) The hole energy $E_{\mathrm{hole}}(\delta)$.
        (b) The $1/\chi$ scaling result of $E_{\mathrm{hole}}(\delta)$.
        (c) The average SC order $\bar\Delta^s$.
        (d) The $1/\chi$ scaling result of $\bar\Delta^s$.
    }
    \label{fig:14}
\end{figure}

\begin{figure}[tbp]
    \centering
    \includegraphics[width=1.0\columnwidth]{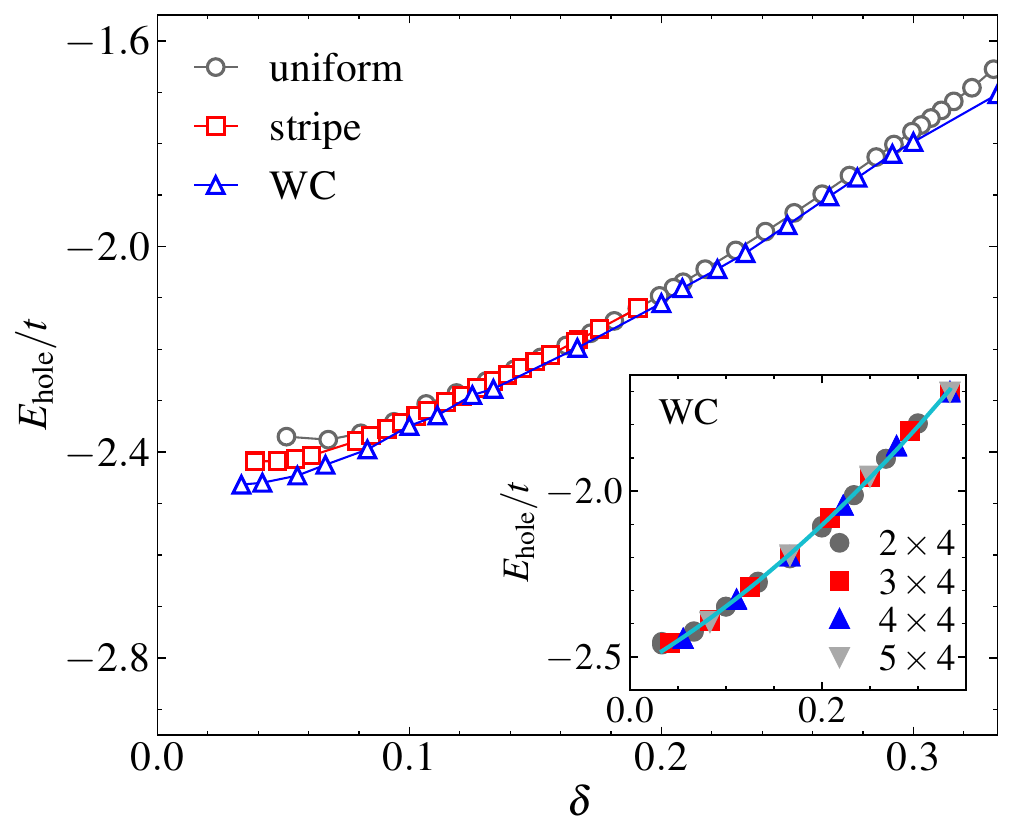} 
    \caption{
        Energies per hole $E_{\mathrm{hole}}$ of various competing states for $t/J = -3.0$ and $D = 14$ as a function of hole doping $\delta$.
        The inset shows WCs on $L_x \times 4$ $(L_x = 2,3,4,5)$ unit cells, which have nearly the same energies per hole.
    }
    \label{fig:15}
\end{figure}

\begin{figure}[tbp]
    \centering
    \includegraphics[width=1.0\columnwidth]{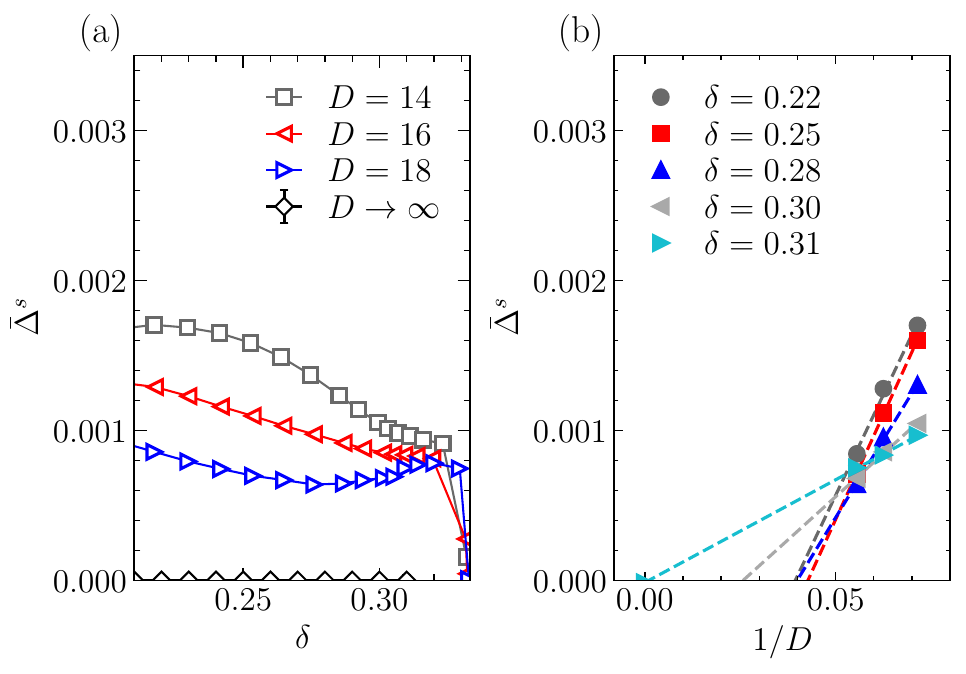} 
    \caption{
        (a) The average SC order $\bar\Delta^s$ for uniform $s$-wave states with $t/J = -3.0$.
        (b) The extrapolation details with $1/D$ for $\bar\Delta^s$ at different hole doping $\delta$.
    }
    \label{fig:16}
\end{figure}

\begin{figure}[tbp]
    \centering
    \includegraphics[width=1.0\columnwidth]{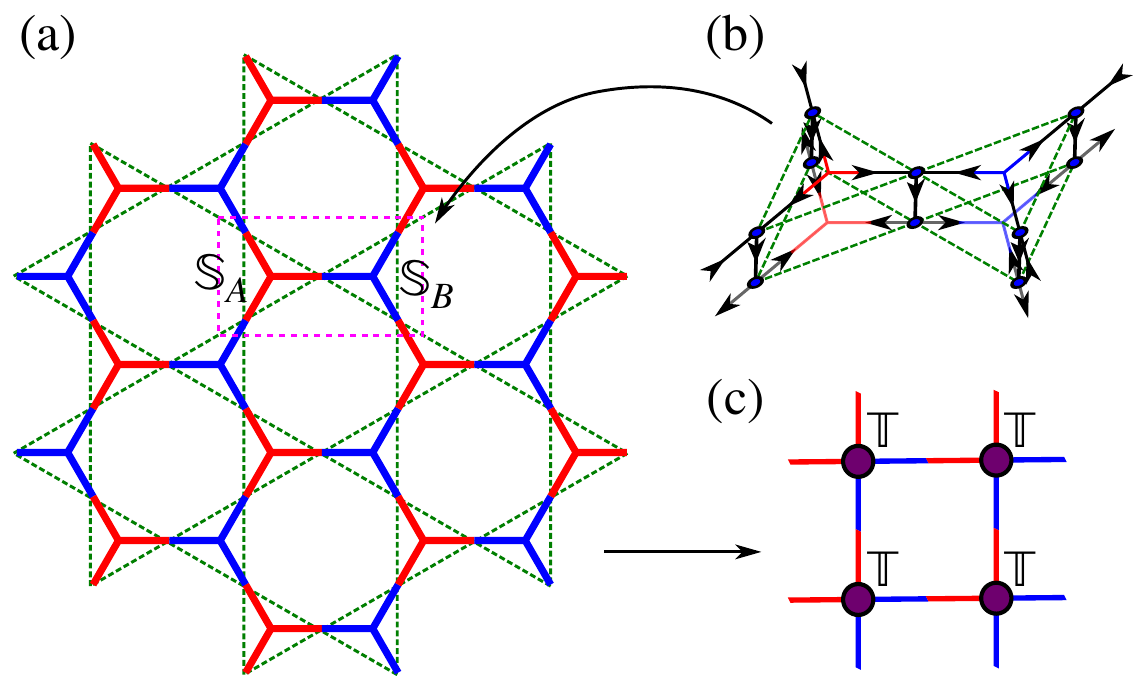} 
    \caption{
        (a) The scalar product $\langle \Psi|\Psi\rangle$ on the Kagome lattice. 
        (b) Details on each unit cell.
        (c) The MPO form on the square lattice.
        For simplicity, we define $\mathbb{T} = \mathcal{C}_v\left(\mathbb{S}_A \mathbb{S}_B\right)$ and represent the scalar product in the MPO form as $\langle \Psi|\Psi\rangle = \mathcal{C}_v\left( \mathbb{S}_A \mathbb{S}_B \cdots \right) = \mathcal{C}_v \left( \mathbb{T} \cdots\right)$.
    }
    \label{fig:17}
\end{figure}

In this section, we report more results on insulating WCs (focusing on charge patterns in other large unit cells), CDW order, average magnetization, and related correlation functions. 
This analysis aims to deepen our understanding of the physical properties of WCs.

Figure \ref{fig:7} shows different charge patterns of WCs in other large $L_x\times 4$ ($L_x \in \{2,4,5\}$) unit cells for various hole doping. 
For WCs in the lightly doped region $\delta < 0.13$ in Figs. \ref{fig:7}(a), \ref{fig:7}(c), and \ref{fig:7}(e), we see that each doped hole forms a localized cluster, whose approximate locations are marked by numbers.
This shows the important role of effective repulsion between doped holes in creating the WC structure.
As hole doping increases to $\delta > 0.13$, the WCs display complex CDW orders in Figs. \ref{fig:7}(b), \ref{fig:7}(d), and \ref{fig:7}(f).
The positions of the maximum local hole density are also marked by numbers, and the number of marked sites is equal to the number of doped holes. 
The clear separation between these maximally doped sites shows the effective repulsive interaction between doped holes for WCs at large hole doping.

Figure \ref{fig:8} plots the CDW order $\Delta_{\mathrm{CDW}}$ of the WCs as a function of $1/D$ in various unit cells.
The CDW order is defined by $\Delta_{\mathrm{CDW}} = \delta_{\mathrm{max}} - \delta_{\mathrm{min}}$, which is the difference between the maximum local hole density ($\delta_{\mathrm{max}}$) and the minimum local hole density ($\delta_{\mathrm{min}}$) at the $L_x\times L_y\times 3$ lattice sites. 
The CDW order of each WC follows the relation $\Delta_{\mathrm{CDW}}(D) = \Delta_0 + \alpha D^{-1}$.
This linear extrapolation with $1/D$ confirms the existence of a consistent CDW order for $\delta < 0.27$.
Moreover, the CDW order decreases with increasing hole doping, suggesting a gradual transition from CDW states to uniform states.

Figure \ref{fig:9} shows the average magnetization $\bar{M}$ of WCs as a function of $1/D$ on various unit cells with light hole doping. 
The average magnetization is computed by $\bar{M} = \sum_{ij\alpha} M_{ij\alpha}/3L_xL_y$, where $M_{ij\alpha} = \sqrt{\langle S^x_{ij\alpha} \rangle^2 + \langle S^y_{ij\alpha} \rangle^2 + \langle S^z_{ij\alpha} \rangle^2}$ is the magnetization at each lattice site $(i,j,\alpha\in\{a,b,c\})$.
We use a quadratic fitting with $1/D$ to extrapolate the average magnetization.
When $D \rightarrow \infty$, each WC shows a reduced magnetization or, in some cases, a zero average magnetization. 
Our results indicate that the insulating WCs are mainly made of spinless holons, consistent with previous DMRG simulations \cite{PhysRevLett.119.067002,peng2021doping}.

Figure \ref{fig:10} plots the single-particle Green's function $G(r)$ for WCs on different unit cells.
The single-particle Green's functions for WCs are given by
\begin{equation}
    G(r) = \frac{1}{L_y} \sum_{y=1}^{L_y}\sum_{\sigma} \left| \langle \hat{c}_{x,y,\sigma}^\dagger \hat{c}_{x+r,y,\sigma}\rangle \right|.
\end{equation}
Here, $r$ is the distance between two sites, including only the $b$ and $c$ sites along the $\vec{v}_1$ direction.
$G(r)$ in semi-logarithmic scale shows $G(r) \sim e^{-r/\xi_G}$.
As hole doping increases, the correlation length $\xi_G$ increases and gradually approaches the correlation length of NFL.

Figure \ref{fig:11} plots the spin-spin correlation functions $F(r)$ for WCs on different unit cells with large hole doping.
The spin-spin correlation functions for WCs are given by
\begin{equation}
    F(r) = \frac{1}{L_y}\sum_{y=1}^{L_y} |\langle \bold{S}_{x,y} \cdot \bold{S}_{x+r,y}\rangle |.
\end{equation}
Here, $F(r)$ on the double logarithmic scale shows $F(r) \sim r^{-K_s}$ for WCs with large hole doping at a short-range distance.
However, $F(r) \rightarrow \frac{1}{L_y}\sum_{y=1}^{L_y} |\langle \bold{S}_{x,y}\rangle \cdot \langle \bold{S}_{x+r,y}\rangle |$ in the long distance, where nonzero local magnetization $\langle \bold{S}_i\rangle$ biases $F(r)$ at finite $D$.
Therefore, we focus only on the short-range spin-spin correlation functions.
The difference between $F(r) \sim r^{-K_s}$ for WC and $F(r) \sim e^{-r/\xi_F}$ for NFL reveals distinct characteristics for these two phases.

\section{Uniform Phase}

We use a linear extrapolation with $1/D$ to confirm the SC order for these uniform states.
Figures \ref{fig:12} show the extrapolation details for the SC order of uniform $s$-wave and PDW states with $D=16,18,20$, respectively.
We find that the SC order of the PDW states exists for $\delta > 0.32$ in the extrapolation. 
However, uniform ``$s$-wave'' states do not have SC order for $D \rightarrow \infty$, and they are NFL for $0.27 < \delta < 0.32$.

Figure \ref{fig:13} shows the SC order of the uniform $d$+$id$-wave states and the extrapolation details with $1/D$.
For all finite $D$ and even $D\rightarrow\infty$, the SC order exists for a wide range of hole doping.
However, chiral $d$+$id$ states are not ground states, as they have higher energies than the NFL and PDW states.

We use the VUMPS algorithm to contract the two-dimensional tensor network and compute related physical quantities.
To check the convergence of physical quantities with the environmental bond dimension $\chi$, we show the hole energy $E_0(\delta)$ and the average SC order $\bar\Delta^s$ for uniform states (NFL and PDW) with $D=16$ and different $\chi$ in Fig. \ref{fig:14}.
Our results indicate a convergence trend in SC order and hole energy when $\chi \gtrsim 4D$.

\section{Results with $t/J=-3.0$}
\label{app:oppot}

To ensure the influence of the sign of $t$ on the insulating WC phase, we investigate properties of the Kagome lattice $t$-$J$ model with $t/J = -3.0$.
Depending on the size of the unit cell and the doping density $\delta$, we also discover uniform states and a series of CDW states.
Similar to results with $t/J = 3.0$, stripe states are found in $L_x\times 1$ unit cells ($L_x \geq 2$), while WCs are found in $L_x\times L_y$ unit cells ($L_x L_y \geq 8$, $L_x \geq 2$, and $L_y \geq 2$).

Figure \ref{fig:15} presents the energy comparison between various competing states for $D = 14$ with $t/J = -3.0$.
Unlike the results with $t/J = 3.0$, we still find WCs near $1/3$ doping with the opposite sign of $t$.
In the doping region $0 < \delta < 1/3$, the WCs have energies lower than those of other competing states.
It also indicates that the emergence of WCs does not depend on the sign of $t$ at low doping.
The result with $D=12$ shows the same trend in the phase transition, confirming the reliable phase diagram.

According to spin-singlet pairing symmetry, uniform states only exhibit the $s$-wave pairing symmetry at $\delta < 1/3$ for finite $D$.
As shown in Fig. \ref{fig:16}, we display the SC order of uniform $s$-wave states and the extrapolation details with $1/D$.
Similar to the results with $t/J = 3.0$, the SC order of the ``$s$-wave'' state also vanishes for $\delta < 0.31$ in the extrapolation as $D\rightarrow \infty$, indicating the NFL with the opposite sign of $t$.

\section{Fermionic Tensor Network}

The fermionic tensor network is an important method for studying electronic systems.
It has various forms, such as the Grassmann tensor product state (GTPS) \cite{gu2010grassmann, PhysRevB.88.115139}, the tensor network with fermionic $\mathbb{Z}_2$-graded vector space \cite{PhysRevB.95.075108, Bultinck_2017}, and the fermionic swap gate approach \cite{PhysRevB.80.165129, PhysRevA.81.010303, PhysRevB.81.165104, PhysRevA.81.052338}. 
Despite their different representations, these fermionic tensor network approaches are essentially equivalent.
In our study, we implement the tensor network based on the fermionic $\mathbb{Z}_2$-graded vector space.

We study fermionic systems on the Kagome lattice by applying the $\mathbb{Z}_2^f$ parity symmetry on the PESS.
In the thermodynamic limit, the tensor network state in the Kagome lattice consists of regularly repeated supercells, which contain the $L_x\times L_y\times 3$ lattice sites in Fig. \ref{fig:1}(b).
We write PESS representation with the $\mathbb{Z}_2^f$ parity symmetry as
\begin{equation}
    |\Psi\rangle = \mathcal{C}_v \left( \prod_{r} \mathcal{S}_{r;A} \mathcal{S}_{r;B} \mathcal{T}^{m_a}_r \mathcal{T}^{m_b}_r \mathcal{T}^{m_c}_r \right), 
\end{equation}
where $r$ is the position. 
For simplicity, considering the state on a $1\times 1$ unit cell in Fig. \ref{fig:1}(c), we have
\begin{equation}
    \begin{aligned}
        \mathcal{S}_A     &= \sum_{a_1b_1c_1} S_{a_1b_1c_1} |a_1)|b_1)|c_1) \in V_{a_1} \otimes V_{b_1} \otimes V_{c_1} , \\
        \mathcal{S}_B     &= \sum_{a_2b_2c_2} S_{a_2b_2c_2} |a_2)|b_2)|c_2) \in V_{a_2} \otimes V_{b_2} \otimes V_{c_2} , \\
        \mathcal{T}^{m_i} &= \sum_{i_1,i_2,m_i}T_{i_1 i_2}^{m_i} |m_i\rangle(i_1|(i_2| \in \mathcal{H}^{m_i} \otimes V_{i_1}^* \otimes V_{i_2}^* , 
    \end{aligned}
\end{equation}
where $i = a, b, c$. 
The supervector space $\mathcal{H}^{m_i}$ is the physical space on the tensor $\mathcal{T}^{m_i}$.
The supervector space $V$ is the virtual index on the tensor $\mathcal{S}_{A(B)}$, while the corresponding supervector space $V^*$ is the virtual index on the tensor $\mathcal{T}^{m_i}$.
Each vector in these supervector spaces follows the relevant concepts introduced in Ref. \cite{PhysRevB.95.075108, Bultinck_2017}.
The contraction map $\mathcal{C}_v$ gives $(\alpha'|\alpha) = \delta_{\alpha\alpha'}$, where $(\alpha'| \in V^*$ and $|\alpha) \in V$.
In general, $\mathcal{T}^{m_i}$ is composed of two virtual indices and one physical index $|m_i\rangle$, which includes one vacuum state $|0\rangle$ with even-parity symmetry and two electronic states $\left|\uparrow\right\rangle$ and $\left|\downarrow\right\rangle$ with odd-parity symmetry. 
$\mathcal{S}_{A( B)}$ has three virtual indices. 
Each fermionic tensor satisfies even parity conservation, such as $|m_i| + |i_1| + |i_2| = 0 \,({\rm mod}\, 2)$ for $\mathcal{T}^{m_i}$ and $|a_{1/2}| + |b_{1/2}| + |c_{1/2}| = 0\, ({\rm mod}\, 2)$ for $\mathcal{S}_{A(B)}$.

\section{Imaginary time Evolution and Contraction Scheme}
\label{sec:SU}

We apply the imaginary time evolution operator $e^{-\tau H}$ on an arbitrary initial state $|\Psi_{{\rm init}}\rangle$ to obtain the ground state $|\Psi\rangle = e^{-\tau H} |\Psi_{{\rm init}}\rangle$ in the limit $\tau \rightarrow \infty$.  
However, the imaginary time operator cannot be built directly.
For two-dimensional systems on the Kagome lattice, we divide the operator into $n$ slices of $e^{-\delta\tau H}$ with $\tau = n\delta\tau$, and apply a Trotter-Suzuki decomposition \cite{Suzuki1976} on each slice
\begin{equation}
    e^{-\delta\tau H} = e^{-\delta\tau H_{\bigtriangleup}} e^{-\delta\tau H_{\bigtriangledown}} + \mathcal{O}(\delta \tau^2),
\end{equation}
where the projector $e^{-\delta\tau H_{\bigtriangleup/\bigtriangledown}}$ corresponds to the upper/lower triangular lattice on the Kagome lattice.

Two main methods are commonly used to update the tensor network state: simple update (SU) \cite{PhysRevX.4.011025, PhysRevLett.98.070201} and full update (FU) \cite{PhysRevLett.101.250602}. 
In this work, we use the SU algorithm to obtain the ground state of the Kagome lattice $t$-$J$ model.
Furthermore, due to the action of time evolution gates at three lattice sites, we use high-order singular value decomposition (HOSVD) to decompose the tensor \cite{PhysRevX.4.011025, PhysRevB.86.045139}. 

After obtaining the ground-state wave function, the important task is to calculate related physical quantities, such as variational energy, magnetization, and charge order.
Several methods have been proposed to contract two-dimensional tensor networks, such as the boundary MPS methods \cite{PhysRevLett.101.250602}, the corner transfer matrix renormalization group (CTMRG) method \cite{PhysRevB.80.094403}, the tensor renormalization group (TRG) method \cite{PhysRevLett.99.120601, PhysRevLett.103.160601}, the VUMPS algorithm \cite{10.21468/SciPostPhysLectNotes.7, PhysRevB.97.045145, PhysRevB.98.235148}, etc. 

Here, we employ the VUMPS algorithm for the contraction of the two-dimensional tensor network.
The algorithm is similar to the boundary MPS method in two-dimensional tensor network contraction.
Moreover, based on the tangent space methods for MPS, the VUMPS algorithm is very efficient in finding the ground state on a one-dimensional and quasi-one-dimensional lattice \cite{PhysRevB.97.045145}. 
As shown in Fig. \ref{fig:17}, we transform the scalar product $\langle \Psi|\Psi\rangle$ from the fermionic PESS on the Kagome lattice to the tensor $\mathbb{T}$ on the square lattice.
Therefore, we can deal with the contraction of the tensor network on the Kagome lattice in the same way as with the square lattice strategy.
More details on the contraction of the tensor network in the square lattice have been discussed in our preview work \cite{PhysRevB.108.035144}.

\end{document}